\documentclass[reprint,
 amsmath,amssymb,
 aps,
pra,
]{revtex4-2}

\usepackage{graphicx}
\usepackage{dcolumn}
\usepackage{bm}

\usepackage{xcolor}

\begin{document}

\preprint{APS/123-QED}

\title{Steady-state phases in long-range measurement-only quantum circuits}

\author{Bihui Zhu}

\affiliation{%
 Homer L. Dodge Department of Physics and Astronomy,
The University of Oklahoma, Norman, Oklahoma 73019, USA\\
 Center for Quantum Research and Technology, The University of Oklahoma, Norman, Oklahoma 73019, USA
}%
 \email{bihui.zhu@ou.edu}

\date{\today}

\begin{abstract}

Measurements can drive quantum many-body systems into nontrivial  steady states and induce interesting dynamical phase transitions, rendering measurement-only quantum circuits a useful platform for exploring quantum many-body phases beyond those of equilibrium Hamiltonian systems. 
Here we study a class of long-range measurement-only quantum circuits with competing two-qubit and three-qubit measurements.  We demonstrate that  these circuits exhibit rich steady-state structure and uncover a strong influence of the measurement range on the resulting phases. In particular, states with symmetry-protected topological (SPT) order can emerge with sufficiently short-range measurements beyond the nearest-neighbor limit. These states feature robust topological edge modes, which can also be detected from circuit dynamics. With longer-range measurements, an extended parameter regime emerges in which conventional order parameters are suppressed while spatial correlations remain nontrivial. Moreover, we show that in this circuit model  sufficiently long-range measurements can produce significant entanglement with scaling beyond an area law despite the absence of any unitary evolution.

\end{abstract}

\maketitle


\section{\label{sec:intro}Introduction}
 Measurements provide a powerful tool for shaping quantum many-body states  beyond  their conventional role as readout operations. Repeated measurements can  alter entanglement structure, generate nontrivial steady states, and drive dynamical phase transitions. A prominent example is the measurement-induced phase transition in monitored quantum dynamics, where the steady-state entanglement changes qualitatively as the measurement rate is varied \cite{liQuantumZenoEffect2018,skinnerMeasurementInducedPhase2019,chanUnitaryprojectiveEntanglement2019,liMeasurementDrivenEntanglementTransition2019,baoTheoryPhaseTransition2020,jianMeasurementInducedCriticality2020,zabaloCriticalProperties2020,gullansDynamicalPurification2020,MIPTdmrg2020}. Recent developments in using measurements as an active element of many-body dynamics have opened a new route to studying phases and critical phenomena out of equilibrium,  where the organizing principles can arise not only from Hamiltonian evolution but also from measurement backaction and measurement outcomes \cite{fisher2023random,noelMeasurementInducedQuantum2022,friedmanMeasurementInducedPhases2023,MIPTscalable2020,lavasani2021measurement,universalityMIPTPRL2020,mblMIPT2020,choiQECPRL2020,qecRandomFanPRB2021,liFisherQECPRB2021,baoSymmetryEnriched2021,MIPTising2021,EEnegativityCriticalityFisher2021,topo2dMonitoredPRL2021,zabaloOperatorScaling2022,mixedStateOrderMeasurementfeedback2023,MIPTexpSC2023,localizationMeas2023,purificatoinUniversalPRX2025,googleExp2023,adaptiveExpfeig2023,EEprepZlatkoPRXqu2024,symmetrytopoMonitoredPRB2026}.

A particularly useful setting is provided by measurement-only quantum circuits, where the quantum dynamics is driven entirely by  measurements rather than unitary gates.
Such circuits have been shown to realize rich dynamical behavior, including entanglement phase transitions, purification transitions, and steady states with nontrivial ordering or topological character \cite{khemaniMeasOnly2021,sangMeasOnlyPRR2021,lavasani2021measurement,langEntanglementTransitionProjective2020,measOnlySpinLiquidVijayPRB2023,loopMeasOnlyPRX2023,measOnlyToric2024,kunoProductionLatticeGauge2023,yuGaplessSymmetryProtectedTopological2025}. In particular, measurement-only dynamics can support symmetry-protected topological features and symmetry-enriched critical behavior, demonstrating that measurement-generated steady states can possess structure beyond  entanglement scaling alone \cite{klockeTopologicalOrderEntanglement2022,yuGaplessSymmetryProtectedTopological2025,SPTMeasOnly2022,yuclusterIsingMeasOnly2025}.

Meanwhile, an important ingredient that can affect quantum many-body behavior is long-range connectivity.
Long-range interactions occur naturally in a variety of quantum simulation platforms, including trapped ions, Rydberg atom arrays, dipolar atom ensembles, and cavity-QED systems, where interactions or effective couplings can extend over many lattice spacings \cite{trappendIonReview2021,browaeysReview2020,dipolarReviewChomaz2023,cavityReview2013,LRreview2023}. 
In Hamiltonian systems, such long-range coupling effects can strongly modify both equilibrium phase structure and dynamical properties, including correlation propagation and entanglement behavior, compared with short-range systems \cite{LRreview2023,nonequiLRreview2024,LR1dReview1dJPA2020,LRIsingEEPRL2012,algebraicAlexey2014,LRIsingKitaevVodola2016,johannesLREE2013,gongKaleidoscope2016,richermeNonlocalPropagationCorrelations2014,LRlightconePRL2012}.  
In monitored quantum circuits, long-range interactions have been shown to alter entanglement dynamics and measurement-induced transitions \cite{yaoLR2022,LRMIPTSciPost2022}.  
More recently, long-range measurements have begun to be explored in measurement-only circuit settings, where they can produce dynamical behavior distinct from  circuits with short-range measurements \cite{LRmocPRB2023,gomez2026MOC}. A natural question is therefore how nonlocality entering through the spatial structure of the measurements affects many-body order when the steady state is generated by measurements rather than relying on unitary evolution.

In this work, we address this question using a class  of measurement-only circuits with competing two- and three-qubit projective measurements. The circuit contains cluster measurements favoring a symmetry-protected topological (SPT) structure and Ising-type measurements whose range is tunable through a control parameter $\alpha$. 
In the large $\alpha$ limit, the model reduces to a nearest-neighbor Ising-cluster measurement circuit, while finite $\alpha$ introduces a nonlocal structure that directly competes with the SPT-generating cluster measurements.

We characterize the steady-state phase structure as the measurement probability and measurement range vary. By examining relevant order parameters, we find that an SPT phase survives weak long-range measurement effects at strong cluster measurement rates, despite being destabilized when the Ising measurements become sufficiently long-ranged.  Strong Ising measurements establish a long-range ordered regime analogous to the ferromagnetic spontaneous-symmetry-breaking (SSB) phase in equilibrium many-body systems, which we refer to as the SSB phase here. 
This SSB region is enhanced as longer-range two-qubit measurements couple more distant sites, but it does not simply replace the destroyed SPT phase. Rather, at larger measurement range, another extended regime emerges in which the usual SPT and SSB order parameters are both suppressed. 
This new regime harbors nontrivial structure beyond a simple disordered steady state. The half-chain entanglement entropy is strongly enhanced there and shows clear growth beyond an area law with system size for sufficiently long-ranged measurements.  Correlation functions further reveal competing spin-spin and string correlations, which can exhibit nearly algebraic decay and symmetry-enriched critical-like behavior in part of this regime.

Purification dynamics and edge responses further reveal the topological character of the SPT regime.
In the parameter regime corresponding to the SPT phase, purification dynamics shows a larger nonvanishing residual entanglement entropy at late times than in the SSB regime, indicating the existence of topological edge states. Additionally, the edge-qubit response to boundary perturbations remains robust for large $\alpha$ and strong cluster measurements.

The rest of the paper is organized as follows. In Sec.~\ref{sec:setup}, we introduce the long-range measurement-only circuit model considered in this work. In Sec.~\ref{subsec:SPT}, we characterize the steady-state phase structure using the corresponding order parameters, purification dynamics, and edge responses. In Sec.~\ref{subsec:EE}, we analyze the half-chain entanglement entropy and its scaling with system size. In Sec.~\ref{sec:corr}, we further study spin-spin and string correlation functions. Finally, in Sec.~\ref{sec:conclusions}, we summarize our results and discuss possible future directions.

\section{\label{sec:setup}Circuit model}
\begin{figure}[h]
\centering
\includegraphics[width=0.35\textwidth]{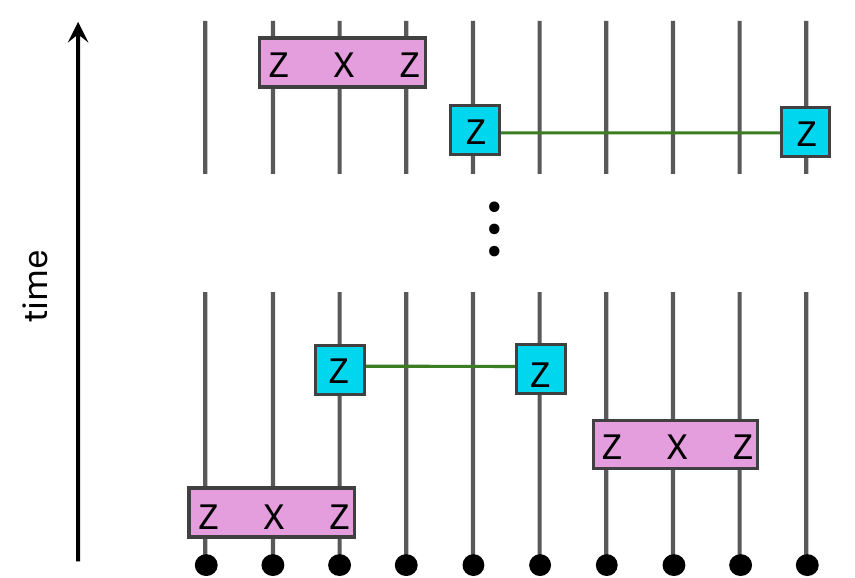}
\caption{Schematic of the long-range measurement-only quantum circuit. The circuit dynamics is driven by two types of measurements, the three-qubit cluster measurements and the ZZ measurements on two qubits. }\label{fig:setup}
\end{figure}

 We study a family of one-dimensional measurement-only circuits as illustrated schematically in Fig.~\ref{fig:setup}. The circuit architecture consists of a set of measurements $\{Z_{i-1}X_iZ_{i+1}, Z_iZ_{i+1}\}$, which preserve a $\mathbb Z_2$ symmetry given by the global term $\prod_{i=1}^L X_i$. At each discrete time step, a measurement is applied randomly: with probability $p_{ZZ}$ a two-qubit measurement $Z_iZ_j$ is applied on a pair of qubits $(i,j)$, drawn from a power-law distribution over the set of all distinct pairs $\mathcal{P}$,
\begin{equation}
    P(i,j)=\mathcal{N}\frac{1}{\vert i-j\vert^{\alpha}},
\end{equation}
 where $\mathcal{N}=(\sum_{(m,n)\in \mathcal{P}} |m-n|^{-\alpha})^{-1}$ is the normalization factor, with open boundary conditions.  With probability $1-p_{ZZ}$, the three-site measurement $Z_{i-1}X_iZ_{i+1}$ is applied, with the center $i$ chosen uniformly from the $L$ qubits on the one-dimensional chain. As such, the two-qubit measurements allow qubits separated by long distances to be coupled together.

Such a circuit model contains no unitary gates; instead, the two types of projective measurements do not commute with each other and drive the quantum dynamics toward a steady state. Since all measurements are Pauli measurements, the circuit consists solely of Clifford operations, which map Pauli strings to Pauli strings. This structure allows the dynamics to be efficiently simulated using the stabilizer formalism based on the Gottesman-Knill theorem \cite{gottesmanClassQuantumErrorcorrecting1996,gottesmanHeisenbergRepresentationQuantum1998,aaronsonImprovedSimulationStabilizer2004}. In this formalism, the many-body state is represented compactly by a set of stabilizer generators, rather than by the full wavefunction in the exponentially large Hilbert space \cite{liMeasurementDrivenEntanglementTransition2019,gullansDynamicalPurification2020,negativityLu2020}. 
In this work, we adopt such an approach to numerically evolve the quantum circuit  and study the steady-state structure generated by the competing measurements.

\section{Phase structure at the steady state}\label{sec:phase}
\subsection{SPT and symmetry-breaking phases}\label{subsec:SPT}
The Hamiltonian counterpart of this circuit model with nearest-neighbor interactions features an SSB phase and an SPT phase in the ground state, when the corresponding two-body Ising term and three-body cluster term dominate, respectively \cite{clusterSPT2015, smacchiaStatisticalMechanicsCluster2011}. 
In the measurement-only circuit model, starting from the initial state $|\psi_0\rangle=\prod_{i=1}^L|+\rangle_i$, with $|+\rangle$  the positive eigenstate of the Pauli $X$ operator, the noncommuting random  measurements generate dynamical evolution of the quantum state. In the steady state, the system exhibits distinct orders depending on the competition between the two types of measurements. In particular, a circuit with only the three-qubit $Z_{i-1}X_{i}Z_{i+1}$ measurements projects the system into the cluster state with SPT order \cite{lavasani2021measurement,son2012topological}; with only the two-qubit measurements, a long-range order emerges in the steady state analogous to the SSB phase in the ground state scenario \cite{sangMeasurementprotectedQuantumPhases2021}. In both cases, the entanglement entropy obeys an area law with the system size. These different steady-state phases can be characterized through the following order parameters: 
\begin{eqnarray}
    O_{\rm SSB}&=&|\langle Z_aZ_b\rangle|, \\
    O_{\rm SPT}&=&|\langle Z_{a-1}Y_a\prod_{a<k<b} X_kY_bZ_{b+1}\rangle|,
\end{eqnarray}
in which we fix $a$ ($b$) to be at a distance $L/4$ from the left (right) edge of the qubit chain to reduce the boundary effect. 

In the quantum circuit setting, the state of the system evolves along different trajectories under random measurements. In the context of measurement-induced phase transitions, the dynamical phase transition is generally not captured by linear observables evaluated on the ensemble-averaged density matrix of the system, but instead depends on trajectory-resolved properties \cite{liMeasurementDrivenEntanglementTransition2019,skinnerMeasurementInducedPhase2019}. Here, a direct average of the signed expectation values of these correlation operators over trajectories can vanish due to cancellations between symmetry-related trajectories. To detect the phase transitions, these order parameters are therefore computed by taking the absolute value of the trajectory-resolved expectation value before averaging over trajectories \cite{morral-yepesDetectingStabilizingMeasurementinduced2023}. 
In this work, for all results shown we average over at least $10^4$ trajectories for each circuit parameter setting considered, evolve circuits for at least $4L^2$ time steps to establish the steady state, and focus on system size $L=128$ unless otherwise specified.

\begin{figure}[h]
\centering
\includegraphics[width=0.45\textwidth]{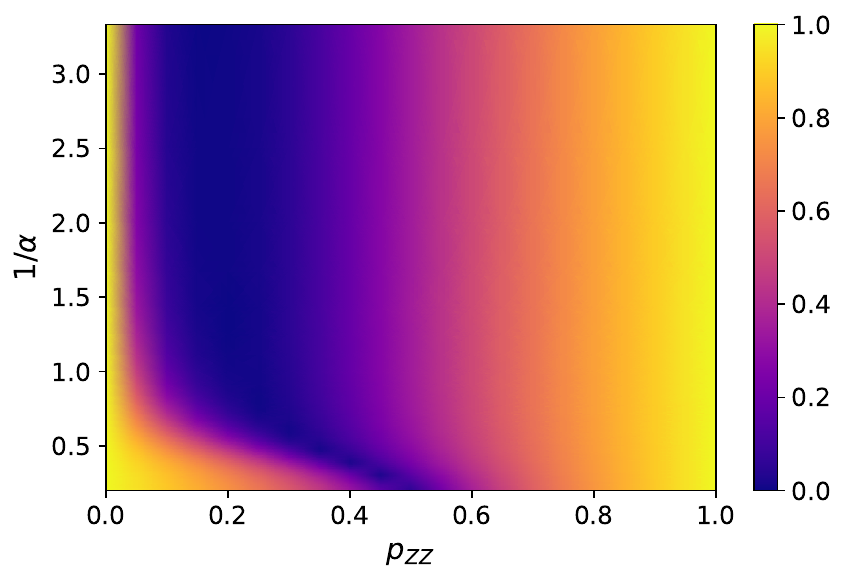}
\caption{Steady-state phase diagram of the long-range measurement-only circuit, characterized by ${\rm max}(O_{\rm SSB},O_{\rm SPT})$ (color bar).}\label{fig:phase}
\end{figure}

With nearest-neighbor-only measurements, the phase transition between the SPT and SSB phases is expected to occur at $p_{ZZ}=p_{ZXZ}=0.5$ \cite{yuGaplessSymmetryProtectedTopological2025}. The presence of long-range coupling affects the competition between these two types of measurements that drive the circuit dynamics and modifies the steady-state phase structure.  
 In Fig. \ref{fig:phase}, we evolve the circuit under the competing measurements and plot the larger of the two order parameters, namely ${\rm max}(O_{\rm SSB}, O_{\rm SPT})$, computed in the steady state and across different measurement probabilities $p_{ZZ}$ and the parameter controlling the range of measurements, $1/\alpha$. For $1/\alpha\ll1$, the  behavior of the system  approximates the case of nearest-neighbor measurements and undergoes a phase transition when $p_{ZZ}\approx p_{ZXZ}$. As $1/\alpha$ increases,  far apart qubits  can be effectively coupled, which enhances the collective order built from the two-qubit $ZZ$ measurements, and the SSB phase region extends to smaller $p_{ZZ}$. Nevertheless, the SPT phase appears to persist at small $1/\alpha$ and sufficiently low  $p_{ZZ}$, before longer-range two-qubit measurements eventually destabilize the SPT order. In Appendix~\ref{app:Stopo}, we also examine the corresponding generalized topological entropy, which serves as an order parameter for the SPT phase in short-range systems and indicates behavior consistent with these findings \cite{lavasani2021measurement,zengTopologicalErrorcorrectingProperties2016a}. Further  finite-size scaling analysis  (see Appendix~\ref{app:finiteL}) 
 suggests that the SPT phase is destroyed for long-range measurements with $\alpha<2$. Meanwhile, longer-range measurements do not continue to establish the SSB order at smaller $p_{ZZ}$, which is interrupted in the presence of frequent cluster measurements; instead, an extended region emerges below moderately large $p_{ZZ}$ where both $O_{\rm SSB}$ and $O_{\rm SPT}$ appear vanishingly small.

\begin{figure}[h]
\centering
\includegraphics[width=0.5\textwidth]{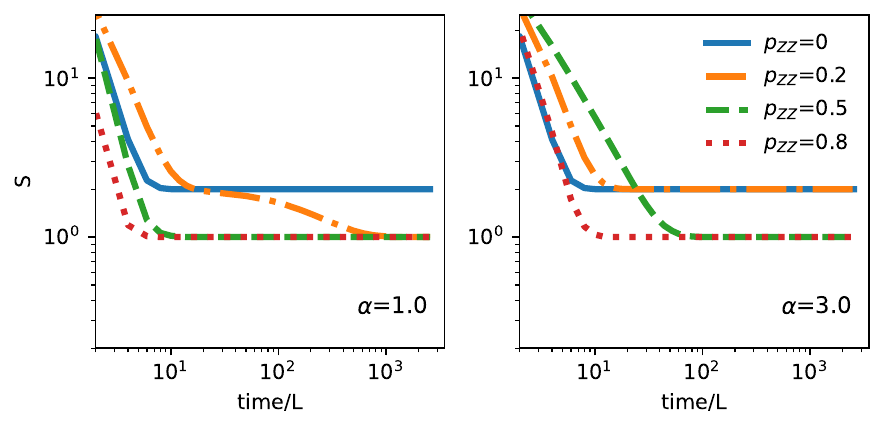}
\caption{Purification dynamics. The time-evolution of the full-system entanglement entropy $S$  is plotted for two different ranges of the $ZZ$ measurements, $\alpha=1$ (left) and $\alpha=3$ (right). Different lines represent the dynamics obtained at various different measurement probabilities $p_{ZZ}$. }\label{fig:pur}
\end{figure}

To analyze the structure of the steady state generated by the circuit dynamics, we also examine the purification dynamics in this circuit model, starting from a maximally mixed initial state $\rho=\mathbb I/2^L$ \cite{gullansDynamicalPurification2020}. In Fig. \ref{fig:pur},  the full system entanglement entropy $S(t)$ is shown for two representative measurement ranges, with $\alpha=1$ and $\alpha=3$ and for different measurement probabilities $p_{ZZ}$. 
With $p_{ZZ}=0$, the steady state reaches the cluster state with SPT order, and the residual entropy $S(t)=2$ at late times indicates two bits of unpurified information, reflecting the existence of four-fold degenerate topological edge states on the open chain \cite{son2012topological,cuiLocalCharacterizationOnedimensional2013}. 
For $\alpha=1$,  with both small and large $p_{ZZ}>0$, the late-time residual entropy is reduced to $S(t)=1$, indicating that the four-fold SPT edge degeneracy is no longer present, with the remaining residual entropy arising from a different, symmetry-breaking origin. In contrast, for $\alpha=3$, the residual entropy remains $S(t)=2$ at late times for small $p_{ZZ}$ within the time window considered, consistent with the presence of edge states associated with the SPT order.

To probe this edge structure more directly at the steady state, we implement an edge-probe protocol. In equilibrium Hamiltonian systems, an SPT-ordered phase on an open chain supports boundary degrees of freedom associated with the nontrivial bulk topology \cite{pollmannSymmetryProtectionTopological2012}. A small boundary field can polarize these edge modes and induce a strong edge response, even when the field remains much smaller than the bulk gap. For the cluster SPT, the open boundary supports effective edge spin-1/2 degrees of freedom, which can be probed through their response to appropriate boundary operators \cite{son2012topological}.
In contrast, in a topologically trivial phase, where no protected boundary degree of freedom is present, the response is expected to scale approximately linearly with the applied field in the weak-field regime, exhibiting  behavior distinct from the SPT case.

\begin{figure}[h]
\centering
\includegraphics[width=0.5\textwidth]{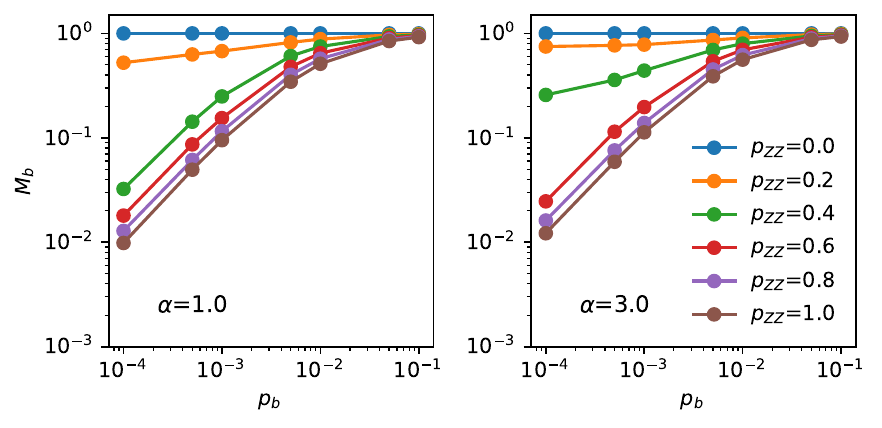}
\caption{Edge responses  at the steady state when a boundary measurement perturbation is applied with probability $p_b$, in a circuit with the two-qubit measurements of $\alpha=1$ (left) and $\alpha=3$ (right). Different colors represent the results obtained at different two-qubit measurement probabilities, ranging from $p_{ZZ}=0$ to $p_{ZZ}=1$.}\label{fig:probe}
\end{figure}

In our circuit setting, we implement the analogous probe dynamically by adding measurements $X_1Z_2$ and $Z_{L-1}X_L$ on the left and right boundary qubits with probability $p_{b}$ \cite{Paszko2024,topodipolarsymPRB2024,preskill2017ph219}. We then characterize the resulting edge polarization at the steady state by 
\begin{eqnarray}
M_b&=&\frac{|\langle X_1Z_2\rangle|+|\langle Z_{L-1}X_L\rangle|}{2}.
\end{eqnarray}
In Fig. \ref{fig:probe}, we show the edge response as a function of the  boundary measurement probability  $p_b$ in different parameter regimes. For large $p_{ZZ}$, $M_b$ decreases rapidly as $p_b$ is reduced, consistent with the system being outside the nontrivial SPT phase when the two-qubit measurements dominate. In contrast, at $p_{ZZ}=0$, $M_b$ stays finite at weak boundary measurement strengths, as expected for the cluster SPT state. For $\alpha=3$ with sufficiently small $p_{ZZ}$, $M_b$ remains robust at weak  $p_b$,  providing further evidence for a nontrivial SPT steady-state regime.

\subsection{Beyond area-law entanglement\label{subsec:EE}}
\begin{figure}[h]
\centering
\includegraphics[width=0.45\textwidth]{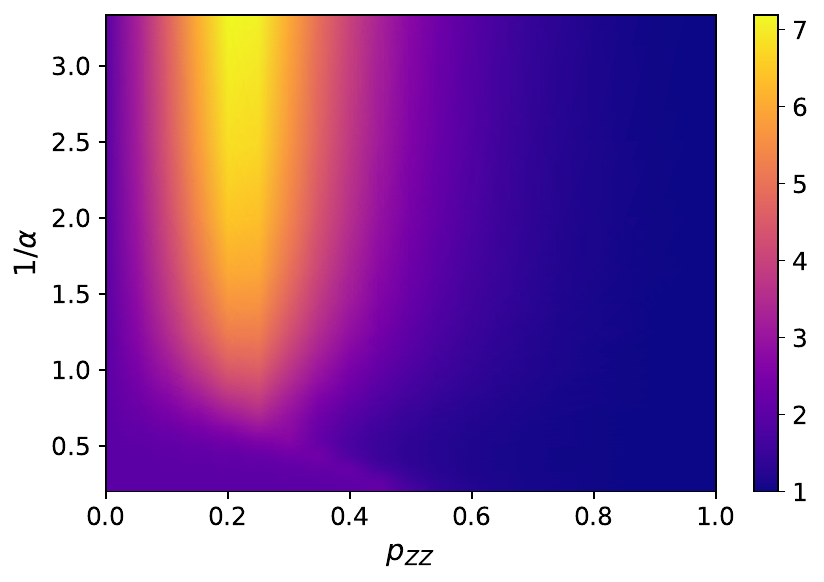}
\caption{Half-chain entanglement entropy at the steady state. The color bar indicates the value of $S_{\rm half}$.  }\label{fig:EEphase}
\end{figure}

In this section, we examine the entanglement generated in the steady state by the circuit dynamics starting from the initial product state. 
Specifically, we compute the half-chain von Neumann entanglement entropy $S_{\rm half}=-{\rm Tr}[\rho_{\rm half}\log \rho_{\rm half}]$, where $\rho_{\rm half}$ is the reduced density matrix of the half-chain subsystem, across different measurement probabilities $p_{ZZ}$ and measurement ranges controlled by $\alpha$.  For the stabilizer circuits considered in this work, this entropy is identical to the R\'enyi entropies as stabilizer states have flat entanglement spectra \cite{stabilizerEE2004}. 
As shown in Fig.~\ref{fig:EEphase}, for short-range measurements, $1/\alpha\ll 1$, and away from the phase boundary, the generated entanglement remains relatively small,  while it becomes strongly enhanced in the intermediate parameter regime and with small $\alpha$. This enhancement is associated with the same part of the phase diagram  in which Fig.~\ref{fig:phase}  indicates neither SPT nor SSB order, suggesting the emergence of a highly entangled steady-state regime.

\begin{figure}[h]
\centering
\includegraphics[width=0.35\textwidth]{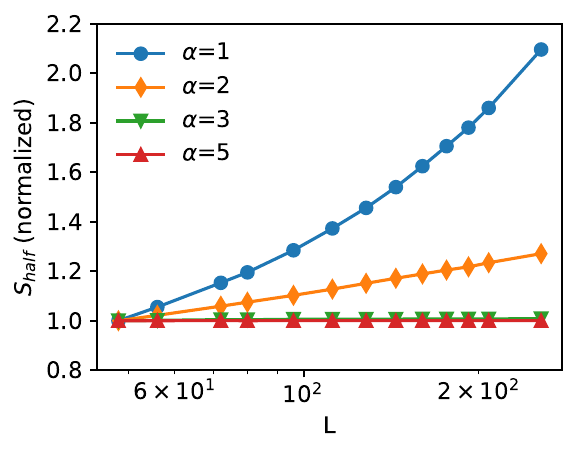}
\caption{Entanglement entropy scaling with system sizes. The half-chain entanglement entropy at the steady-state is computed at fixed  $p_{ZZ}=0.3$ for various $\alpha$'s. For each $\alpha$, $S_{\rm half}(L)$ normalized by the corresponding value at the smallest $L$ is plotted. Here $L=48$ is used. }\label{fig:EEscaling}
\end{figure}

To further check the entanglement behavior in the intermediate parameter regime,  we compute the steady-state $S_{\rm half}$ as a function of system size $L$ at fixed $p_{ZZ}$ and for different measurement ranges controlled by $\alpha$. 
For comparing the entanglement scaling with $L$ across different $\alpha$'s, we normalize the steady-state $S_{\rm half}$ at each parameter $\alpha$ by its value obtained for the smallest system size $L$ considered in Fig.~\ref{fig:EEscaling}.
As this figure shows, the measurement range induces qualitatively different behaviors in steady-state entanglement scaling. 
For long-range measurements with $\alpha\lesssim 2$ considered here, a clear growth in the entanglement beyond an area law is found. In particular, at $\alpha=2$, the half-chain entanglement entropy exhibits critical-like logarithmic scaling. As shown for $\alpha\geq 3$, the entanglement growth remains close to an area law. These results are consistent with the observation of a strongly increased entanglement entropy in the intermediate region in Fig.~\ref{fig:EEphase}. We note that in systems with power-law interacting Hamiltonians and long-range unitary gates, similar strong modifications of the  entanglement properties have been suggested \cite{LRIsingEEPRL2012, fermionLR2022,arealawLRgong2017,algebraicAlexey2014,yaoLR2022}, while here the entanglement arises solely from projective measurements rather than unitary time evolution.

\section{Correlations and symmetry-enriched behavior\label{sec:corr}}
In this section, we take a closer look at the intermediate regime    revealed in Figs. \ref{fig:phase} and  \ref{fig:EEphase}, where both the SSB and the SPT order parameters, $O_{\rm SSB}$ and ${O_{\rm SPT}}$, appear suppressed. To this end, we compute the connected spin-spin correlation function $C_{ZZ}(r)$ and the nonlocal string correlation function $C_{\rm SPT}(r)$ at the steady state as a function of the site distance $r$, defined as
\begin{eqnarray}
    C_{ZZ}(r)&=&|\langle Z_iZ_{i+r}\rangle-\langle Z_i\rangle\langle Z_{i+r}\rangle|, \\
    C_{\rm SPT}(r=|j-i|)&=&|\langle Z_{i-1}Y_i\prod_{k=i+1}^{j-1}X_k Y_jZ_{j+1}\rangle|.
\end{eqnarray}

\begin{figure}[h]
\centering
\includegraphics[width=0.5\textwidth]{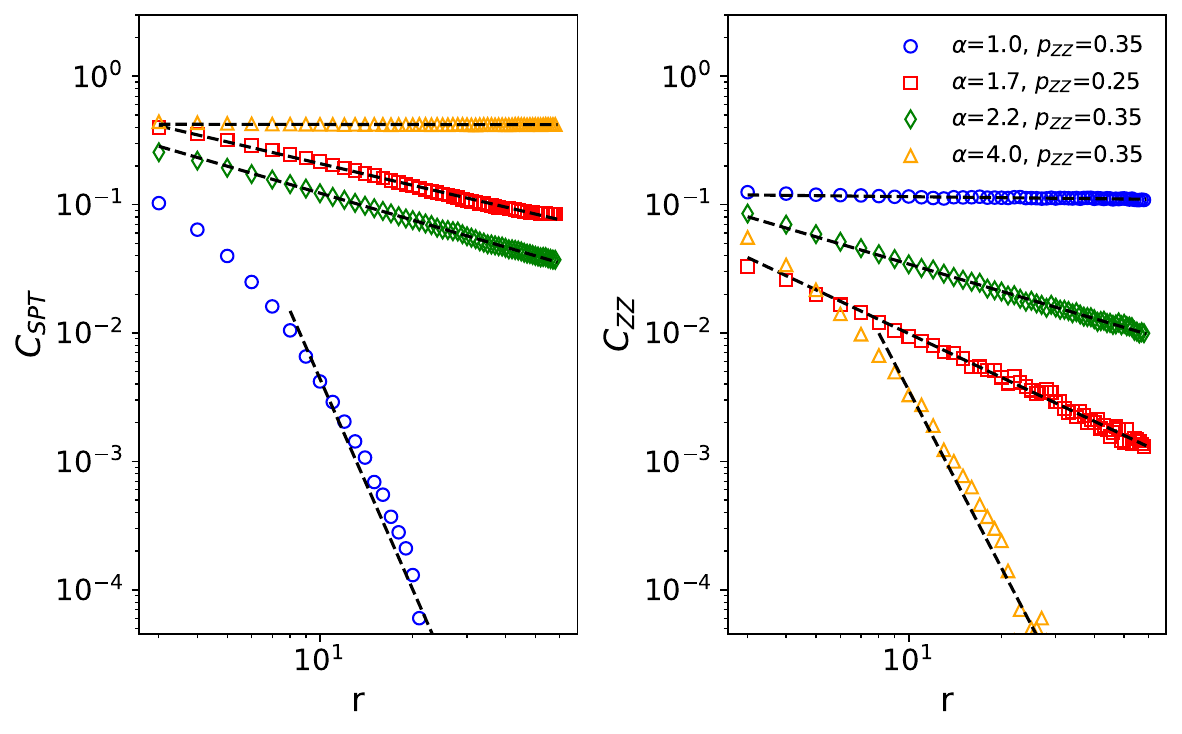}
\caption{Decay of string correlations $C_{\rm SPT}$ (left) and connected spin-spin correlations $C_{ZZ}$ (right), plotted in double logarithmic scale. }\label{fig:corr}
\end{figure}

For a range of moderate $\alpha$ and $p_{ZZ}$ values, we find both the spin correlations $C_{ZZ}(r)$ and string correlations $C_{\rm SPT}(r)$ display approximately power-law decay with distance $r$, reminiscent of the  behavior of critical states in short-range interacting systems. 
Interestingly, in part of this intermediate region,  $C_{\rm SPT}(r)$ decays more slowly than $C_{ZZ}(r)$, suggesting a symmetry-enriched steady-state regime of this circuit \cite{scaffidiGaplessSymmetryProtectedTopological2017,verresenGaplessTopologicalPhases2021}.
Fig.~\ref{fig:corr} shows the decay of correlations at representative parameter points, with power-law fits of the form $C(r)=Ar^{-\Delta}$ for each case also plotted for reference. With short-range measurements and intermediate $p_{ZZ}=0.35$ (orange), the string correlations appear constant across large distances, while the connected spin-spin correlations $C_{ZZ}(r)$ decrease rapidly with distance, consistent with the system being in the SPT phase as discussed in Sec.~\ref{subsec:SPT}.  At intermediate $\alpha$ and $p_{ZZ}$ considered here, both correlations exhibit algebraic decay over a broad range of distances, illustrating a nontrivial interplay between long-range interactions and topology.
In particular, for the results colored in red, $\Delta_{\rm SPT}\approx 0.56$ and   $\Delta_{ZZ}\approx 1.13$ respectively, 
indicating that the string correlations dominate over the two-body spin-spin correlations at large distances. This suggests a scenario where the SPT character remains visible in the long-distance correlations despite the apparent suppression of the string order parameter. 
The results colored in green demonstrate the scenario in which the two types of correlations exhibit comparable long-distance scaling, with $\Delta_{\rm SPT}\approx 0.7$ and   $\Delta_{ZZ}\approx 0.7$. Related behaviors in ground-state phases of Hamiltonian systems  were also found in a recent study, where the power-law interactions introduce 
anomalous effects absent in short-range interacting systems \cite{algebraicSPT2025}.    
For the small  $\alpha$ (blue) shown in the figure, the string correlation  decays exponentially, suggesting the absence of the SPT order, while the connected spin-spin correlation displays only a very weak decay over the distances accessed, with $\Delta_{ZZ}\approx 0.03$, suggesting a quasi-long-range order resulting from sufficiently long-range measurements. 

\section{\label{sec:conclusions} Summary and outlook}
In this work, we studied a class of long-range measurement-only circuits constructed from competing  two-qubit Ising measurements and three-qubit cluster measurements. The Ising measurements involve two qubits separated at distances beyond nearest neighbors, with the effective range tunable through $\alpha$. Using order parameters derived from the nonlocal string correlations and ferromagnetic correlations used to detect SPT and SSB orders in equilibrium phases and adapted to the nonequilibrium circuit settings, we showed that the steady states in these circuits can display distinct measurement-induced orders depending on the strength of the measurements and the measurement range. In particular, we demonstrated the persistence of SPT order beyond the nearest-neighbor limit, whose presence also manifests through a larger residue entropy in the purification dynamics and robust edge responses. We also found that long-range measurements introduce an extended region in parameter space which does not feature pronounced order parameters while exhibiting rich structure through spatial string and connected spin-spin correlations. We further analyzed the quantum entanglement generated at the steady states, and the modification from long-range effects is found to be strongest in the intermediate parameter regime, with clear beyond-area-law entanglement at small $\alpha$.

Our results illustrate that spatially structured measurements may provide a useful tool for engineering quantum many-body states with nontrivial orders and for studying nonequilibrium phases and critical-like phenomena. The effect of long-range measurements has a significant impact on phase organization beyond simply destroying one phase while stabilizing the other. An interesting direction for future work concerns a more detailed analysis of the phase boundaries and universality classes of relevant phase transitions, especially of the properties in the intermediate $\alpha$ regime. 
Furthermore, the circuit setup considered here realizes a scenario that is related to a ferromagnetic cluster-Ising Hamiltonian counterpart, where long-range Ising couplings are expected to have a strong effect. Another possible direction would be to explore setups where the measurement protocols impose antiferromagnetic parity constraints, for example through suitable postselection of measurement outcomes. In long-range settings, such constraints can become mutually frustrated, potentially leading to a different interplay between long-range effects and topological structure and giving rise to new many-body behaviors. 
The family of circuits considered here provides one example of long-range circuits with nontrivially ordered phases,  and it would be interesting to extend the study to other types of phases, including those that may emerge in higher-dimensional circuits.

\begin{acknowledgments}
We thank the Kavli Institute for Theoretical Physics (KITP) for hospitality while part of this work was completed. We acknowledge support from   the National Science Foundation through
grants LEAPS-MPS 2317030 and CAREER Award 2443398. This work used Anvil at Purdue University through allocation PHY260116  from the Advanced Cyberinfrastructure Coordination Ecosystem: Services \& Support (ACCESS) program, which is supported by U.S. National Science Foundation grants \#2138259, \#2138286, \#2138307, \#2137603, and \#2138296 \cite{Boerner2023ACCESS}.

\end{acknowledgments}

\appendix
\begin{figure}[h]
\centering
\includegraphics[width=0.4\textwidth]{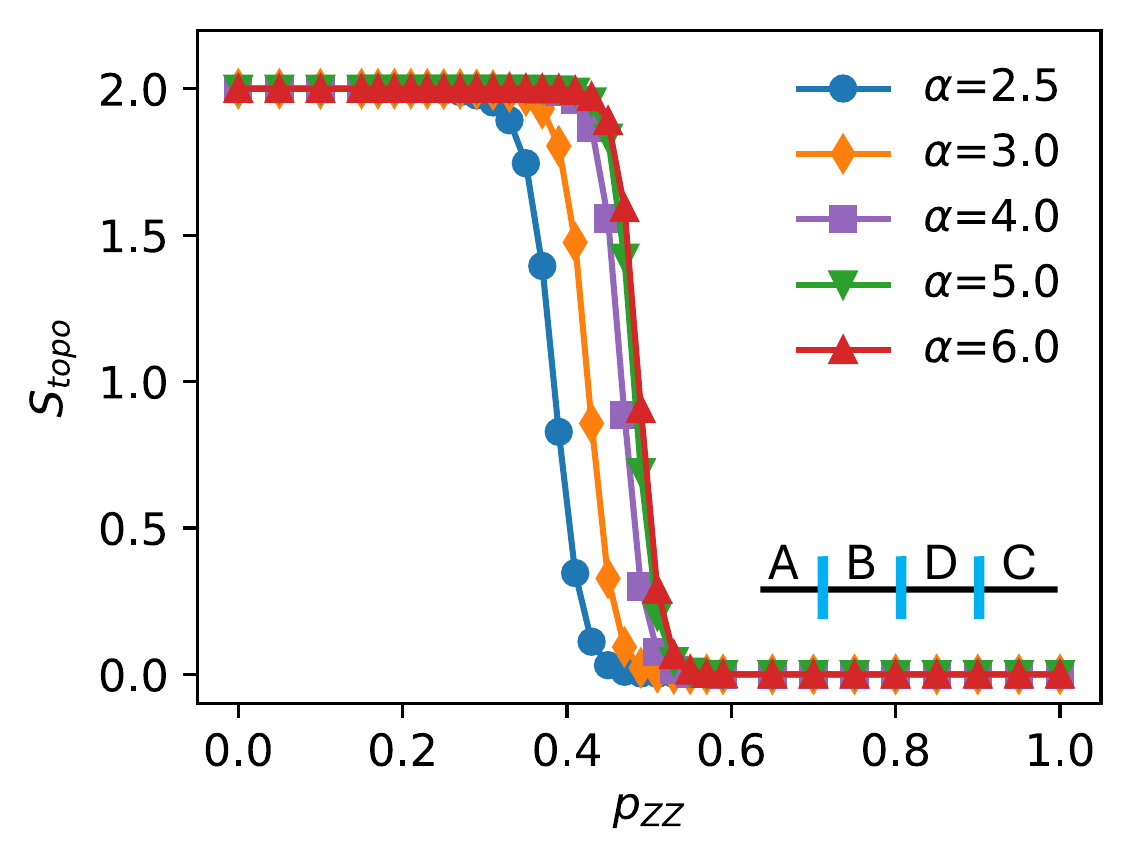}
\caption{Generalized topological entanglement entropy $S_{\rm topo}$. The inset depicts the geometry used in the definition of $S_{\rm topo}$, where the chain is divided into four equal regions.  }\label{fig:Stopo}
\end{figure}
\section{Generalized topological entanglement entropy}\label{app:Stopo}
In this section, we investigate additional probes for the SPT order. To this end, we consider a generalized topological entanglement entropy, which has been used as an order parameter for the SPT phase in gapped ground states  and in nearest-neighbor coupled circuits \cite{zengTopologicalErrorcorrectingProperties2016a,StopoPRB2017,lavasani2021measurement,klockeTopologicalOrderEntanglement2022}. For a one dimensional chain as in our circuit model,   partitioning the system into four regions as depicted in Fig.~\ref{fig:Stopo}, the generalized topological entanglement entropy $S_{\rm topo}$ is defined as:
\begin{equation}
    S_{\rm topo}=S_{AB}+S_{BC}-S_{B}-S_{ABC},
\end{equation}
where $S_{AB}$ denotes the entanglement entropy for the subsystem $AB$.  For gapped Hamiltonian systems, $S_{\rm topo}$ is quantized in the thermodynamic limit and its nonzero value is a signature of SPT order. For the cluster state in particular, $S_{\rm topo}=2$. In Fig.~\ref{fig:Stopo}, we compute $S_{\rm topo}$ as the two-qubit measurement probability varies for several different $\alpha$'s. As sufficiently long-range couplings can contribute additional nonlocal entanglement, we restrict here to  moderately large $\alpha$. As  Fig.~\ref{fig:Stopo} indicates, for small $p_{ZZ}$, $S_{\rm topo}$ stays at $2$, and quickly drops to $0$ above a certain threshold of $p_{ZZ}$, reflecting a transition from a nontrivial SPT order to a trivial one. For large $\alpha$, this threshold is around $0.5$,  and is nearly identical for $\alpha=5$ and $6$, suggesting behavior close to the nearest-neighbor cases. Similar behaviors are observed for smaller $\alpha>2$, while the threshold is shifted toward smaller $p_{ZZ}$ as $\alpha$ is reduced, consistent with the findings revealed from $O_{SPT}$ and $O_{SSB}$ in the main text.

\section{Additional results for different system sizes} \label{app:finiteL}
\begin{figure*}
\centering
\includegraphics[width=0.98\textwidth]{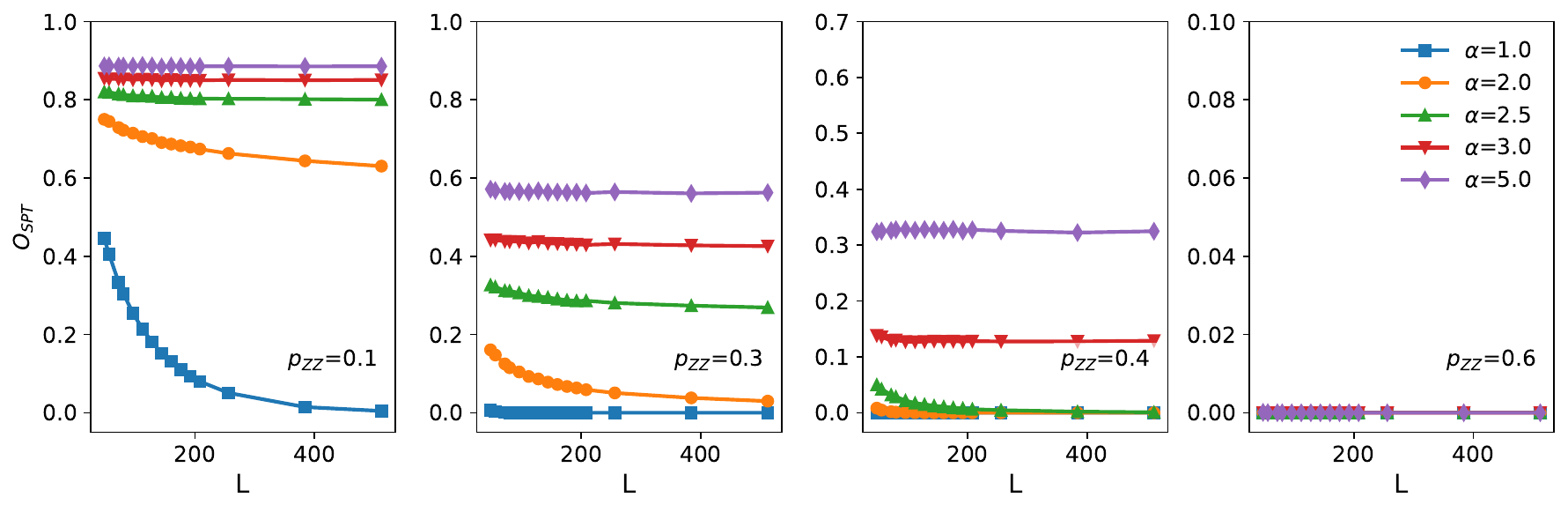}
\caption{String order parameter $O_{\rm SPT}$ at the steady state as the system size $L$ is varied. }\label{fig:scanL-Ospt}
\end{figure*}

\begin{figure*}
\centering
\includegraphics[width=0.95\textwidth]{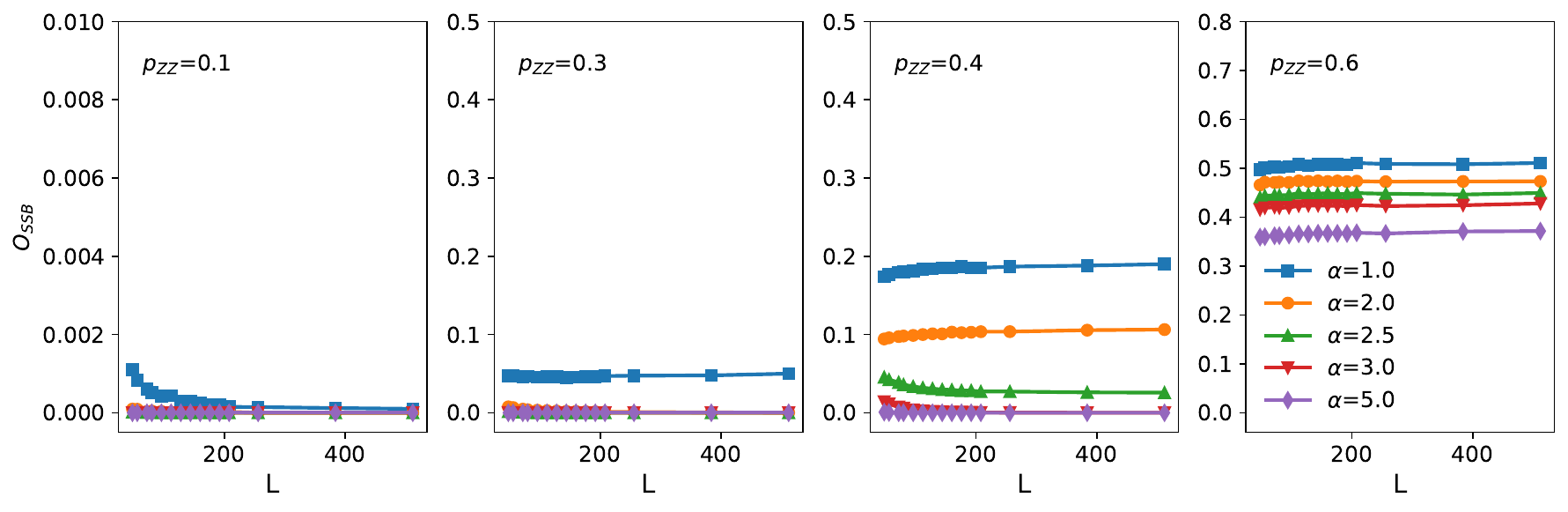}
\caption{SSB order parameter $O_{\rm SSB}$ at the steady state as the system size $L$ is varied. }\label{fig:scanL-Ossb}
\end{figure*}

In Sec.~\ref{sec:phase}, we presented the results obtained from numerically simulating circuits of size $L=128$. Here, we extend the analysis of the order parameters for a range of different system sizes to account for finite-size effects.  In each panel of Fig.~\ref{fig:scanL-Ospt} and Fig.~\ref{fig:scanL-Ossb}, we compute $O_{\rm SPT}$ and $O_{\rm SSB}$, respectively, for representative $\alpha$ values, from very long-range to approximately short-range measurement regimes,  and each panel corresponds to a fixed two-qubit measurement probability, ranging from small to large $p_{ZZ}$ (left to right). As these figures show, for moderately large $\alpha$, the SPT order parameter remains finite within the  accessible system sizes when $p_{ZZ}$ is sufficiently small, while for very small $\alpha$ such as $\alpha=1$, the SSB order parameter remains suppressed at relatively low $p_{ZZ}$. These results provide further evidence for  the phase structure discussed in the main text.

\bibliography{refs/addition, refs/SPT, refs/MIPT-past, refs/LRham}

@article{LRIsingEEPRL2012,
  title = {Entanglement Entropy for the Long-Range Ising Chain in a Transverse Field},
  author = {Koffel, Thomas and Lewenstein, M. and Tagliacozzo, Luca},
  journal = {Phys. Rev. Lett.},
  volume = {109},
  issue = {26},
  pages = {267203},
  numpages = {5},
  year = {2012},
  month = {Dec},
  publisher = {American Physical Society},
  doi = {10.1103/PhysRevLett.109.267203},
  url = {https://link.aps.org/doi/10.1103/PhysRevLett.109.267203}
}

@article{arealawLRgong2017,
  title = {Entanglement Area Laws for Long-Range Interacting Systems},
  author = {Gong, Zhe-Xuan and Foss-Feig, Michael and Brand\~ao, Fernando G. S. L. and Gorshkov, Alexey V.},
  journal = {Phys. Rev. Lett.},
  volume = {119},
  issue = {5},
  pages = {050501},
  numpages = {6},
  year = {2017},
  month = {Jul},
  publisher = {American Physical Society},
  doi = {10.1103/PhysRevLett.119.050501},
  url = {https://link.aps.org/doi/10.1103/PhysRevLett.119.050501}
}

@article{algebraicAlexey2014,
  title = {Kitaev Chains with Long-Range Pairing},
  author = {Vodola, Davide and Lepori, Luca and Ercolessi, Elisa and Gorshkov, Alexey V. and Pupillo, Guido},
  journal = {Phys. Rev. Lett.},
  volume = {113},
  issue = {15},
  pages = {156402},
  numpages = {5},
  year = {2014},
  month = {Oct},
  publisher = {American Physical Society},
  doi = {10.1103/PhysRevLett.113.156402},
  url = {https://link.aps.org/doi/10.1103/PhysRevLett.113.156402}
}

@article{nonequiLRreview2024,
title = {Out-of-equilibrium dynamics of quantum many-body systems with long-range interactions},
journal = {Physics Reports},
volume = {1074},
pages = {1-92},
year = {2024},
issn = {0370-1573},
doi = {https://doi.org/10.1016/j.physrep.2024.04.005},
url = {https://www.sciencedirect.com/science/article/pii/S0370157324001406},
author = {Nicolò Defenu and Alessio Lerose and Silvia Pappalardi},
}

@article{LRreview2023,
  title = {Long-range interacting quantum systems},
  author = {Defenu, Nicol\`o and Donner, Tobias and Macr\`{\i}, Tommaso and Pagano, Guido and Ruffo, Stefano and Trombettoni, Andrea},
  journal = {Rev. Mod. Phys.},
  volume = {95},
  issue = {3},
  pages = {035002},
  numpages = {70},
  year = {2023},
  month = {Aug},
  publisher = {American Physical Society},
  doi = {10.1103/RevModPhys.95.035002},
  url = {https://link.aps.org/doi/10.1103/RevModPhys.95.035002}
}

@article{LR1dReview1dJPA2020,
  title={One-dimensional quantum many body systems with long-range interactions},
  author={Maity, Somnath and Bhattacharya, Utso and Dutta, Amit},
  journal={Journal of Physics A: Mathematical and Theoretical},
  volume={53},
  number={1},
  pages={013001},
  year={2020},
  publisher={IOP Publishing}
}

@article{LRIsingKitaevVodola2016,
  title={Long-range {I}sing and {K}itaev models: phases, correlations and edge modes},
  author={Vodola, Davide and Lepori, Luca and Ercolessi, Elisa and Pupillo, Guido},
  journal={New Journal of Physics},
  volume={18},
  number={1},
  pages={015001},
  year={2016},
  publisher={IOP Publishing}
}

@article{johannesLREE2013,
  title = {Entanglement Growth in Quench Dynamics with Variable Range Interactions},
  author = {Schachenmayer, J. and Lanyon, B. P. and Roos, C. F. and Daley, A. J.},
  journal = {Phys. Rev. X},
  volume = {3},
  issue = {3},
  pages = {031015},
  numpages = {16},
  year = {2013},
  month = {Sep},
  publisher = {American Physical Society},
  doi = {10.1103/PhysRevX.3.031015},
  url = {https://link.aps.org/doi/10.1103/PhysRevX.3.031015}
}

@article{gongKaleidoscope2016,
  title = {Kaleidoscope of quantum phases in a long-range interacting spin-1 chain},
  author = {Gong, Z.-X. and Maghrebi, M. F. and Hu, A. and Foss-Feig, M. and Richerme, P. and Monroe, C. and Gorshkov, A. V.},
  journal = {Phys. Rev. B},
  volume = {93},
  issue = {20},
  pages = {205115},
  numpages = {12},
  year = {2016},
  month = {May},
  publisher = {American Physical Society},
  doi = {10.1103/PhysRevB.93.205115},
  url = {https://link.aps.org/doi/10.1103/PhysRevB.93.205115}
}

@article{richermeNonlocalPropagationCorrelations2014,
  title = {Non-Local Propagation of Correlations in Quantum Systems with Long-Range Interactions},
  author = {Richerme, Philip and Gong, Zhe-Xuan and Lee, Aaron and Senko, Crystal and Smith, Jacob and {Foss-Feig}, Michael and Michalakis, Spyridon and Gorshkov, Alexey V. and Monroe, Christopher},
  year = 2014,
  month = jul,
  journal = {Nature},
  volume = {511},
  number = {7508},
  pages = {198--201},
  publisher = {Nature Publishing Group},
  issn = {1476-4687},
  doi = {10.1038/nature13450},
  copyright = {2014 Springer Nature Limited},
  langid = {english}
}

@article{LRlightconePRL2012,
  title = {Spread of Correlations in Long-Range Interacting Quantum Systems},
  author = {Hauke, P. and Tagliacozzo, L.},
  journal = {Phys. Rev. Lett.},
  volume = {111},
  issue = {20},
  pages = {207202},
  numpages = {6},
  year = {2013},
  month = {Nov},
  publisher = {American Physical Society},
  doi = {10.1103/PhysRevLett.111.207202},
  url = {https://link.aps.org/doi/10.1103/PhysRevLett.111.207202}
}

@article{trappendIonReview2021,
  title = {Programmable quantum simulations of spin systems with trapped ions},
  author = {Monroe, C. and Campbell, W. C. and Duan, L.-M. and Gong, Z.-X. and Gorshkov, A. V. and Hess, P. W. and Islam, R. and Kim, K. and Linke, N. M. and Pagano, G. and Richerme, P. and Senko, C. and Yao, N. Y.},
  journal = {Rev. Mod. Phys.},
  volume = {93},
  issue = {2},
  pages = {025001},
  numpages = {57},
  year = {2021},
  month = {Apr},
  publisher = {American Physical Society},
  doi = {10.1103/RevModPhys.93.025001},
  url = {https://link.aps.org/doi/10.1103/RevModPhys.93.025001}
}

@article{browaeysReview2020,
  title={Many-body physics with individually controlled {R}ydberg atoms},
  author={Browaeys, Antoine and Lahaye, Thierry},
  journal={Nature Physics},
  volume={16},
  number={2},
  pages={132--142},
  year={2020},
  publisher={Nature Publishing Group UK London}
}

@article{dipolarReviewChomaz2023,
  title={Dipolar physics: a review of experiments with magnetic quantum gases},
  author={Chomaz, Lauriane and Ferrier-Barbut, Igor and Ferlaino, Francesca and Laburthe-Tolra, Bruno and Lev, Benjamin L and Pfau, Tilman},
  journal={Reports on Progress in Physics},
  volume={86},
  number={2},
  pages={026401},
  year={2023},
  publisher={IOP Publishing}
}

@article{cavityReview2013,
  title = {Cold atoms in cavity-generated dynamical optical potentials},
  author = {Ritsch, Helmut and Domokos, Peter and Brennecke, Ferdinand and Esslinger, Tilman},
  journal = {Rev. Mod. Phys.},
  volume = {85},
  issue = {2},
  pages = {553--601},
  numpages = {0},
  year = {2013},
  month = {Apr},
  publisher = {American Physical Society},
  doi = {10.1103/RevModPhys.85.553},
  url = {https://link.aps.org/doi/10.1103/RevModPhys.85.553}
}

@article{yaoLR2022,
  title = {Measurement-Induced Transition in Long-Range Interacting Quantum Circuits},
  author = {Block, Maxwell and Bao, Yimu and Choi, Soonwon and Altman, Ehud and Yao, Norman Y.},
  journal = {Phys. Rev. Lett.},
  volume = {128},
  issue = {1},
  pages = {010604},
  numpages = {7},
  year = {2022},
  month = {Jan},
  publisher = {American Physical Society},
  doi = {10.1103/PhysRevLett.128.010604},
  url = {https://link.aps.org/doi/10.1103/PhysRevLett.128.010604}
}

@article{fermionLR2022,
  title = {Fate of Measurement-Induced Phase Transition in Long-Range Interactions},
  author = {Minato, Takaaki and Sugimoto, Koudai and Kuwahara, Tomotaka and Saito, Keiji},
  journal = {Phys. Rev. Lett.},
  volume = {128},
  issue = {1},
  pages = {010603},
  numpages = {7},
  year = {2022},
  month = {Jan},
  publisher = {American Physical Society},
  doi = {10.1103/PhysRevLett.128.010603},
  url = {https://link.aps.org/doi/10.1103/PhysRevLett.128.010603}
}

@article{LRMIPTSciPost2022,
  title={Measurement-induced criticality in extended and long-range unitary circuits},
  author={Sharma, Shraddha and Turkeshi, Xhek and Fazio, Rosario and Dalmonte, Marcello},
  journal={SciPost Physics Core},
  volume={5},
  number={2},
  pages={023},
  year={2022}
}

@article{gomez2026MOC,
  title={Entanglement and information scrambling in long-range measurement-only circuits},
  author={Gomez, Abigail McClain and Abney-McPeek, Fiona and Hu, Hong-Ye and Yelin, Susanne F and Da{\u{g}}, Ceren B},
  journal={arXiv preprint arXiv:2604.22022},
  year={2026}
}

@article{LRmocPRB2023,
  title = {Phase transition and evidence of fast-scrambling phase in measurement-only quantum circuits},
  author = {Kuno, Yoshihito and Orito, Takahiro and Ichinose, Ikuo},
  journal = {Phys. Rev. B},
  volume = {108},
  issue = {9},
  pages = {094104},
  numpages = {12},
  year = {2023},
  month = {Sep},
  publisher = {American Physical Society},
  doi = {10.1103/PhysRevB.108.094104},
  url = {https://link.aps.org/doi/10.1103/PhysRevB.108.094104}
}

@article{baoTheoryPhaseTransition2020,
  title = {Theory of the phase transition in random unitary circuits with measurements},
  author = {Bao, Yimu and Choi, Soonwon and Altman, Ehud},
  journal = {Phys. Rev. B},
  volume = {101},
  issue = {10},
  pages = {104301},
  numpages = {26},
  year = {2020},
  month = {Mar},
  publisher = {American Physical Society},
  doi = {10.1103/PhysRevB.101.104301},
  url = {https://link.aps.org/doi/10.1103/PhysRevB.101.104301}
}

@article{skinnerMeasurementInducedPhase2019,
  title = {Measurement-Induced Phase Transitions in the Dynamics of Entanglement},
  author = {Skinner, Brian and Ruhman, Jonathan and Nahum, Adam},
  journal = {Phys. Rev. X},
  volume = {9},
  issue = {3},
  pages = {031009},
  numpages = {21},
  year = {2019},
  month = {Jul},
  publisher = {American Physical Society},
  doi = {10.1103/PhysRevX.9.031009},
  url = {https://link.aps.org/doi/10.1103/PhysRevX.9.031009}
}

@article{chanUnitaryprojectiveEntanglement2019,
  title = {Unitary-projective entanglement dynamics},
  author = {Chan, Amos and Nandkishore, Rahul M. and Pretko, Michael and Smith, Graeme},
  journal = {Phys. Rev. B},
  volume = {99},
  issue = {22},
  pages = {224307},
  numpages = {16},
  year = {2019},
  month = {Jun},
  publisher = {American Physical Society},
  doi = {10.1103/PhysRevB.99.224307},
  url = {https://link.aps.org/doi/10.1103/PhysRevB.99.224307}
}

@article{liMeasurementDrivenEntanglementTransition2019,
  title = {Measurement-driven entanglement transition in hybrid quantum circuits},
  author = {Li, Yaodong and Chen, Xiao and Fisher, Matthew P. A.},
  journal = {Phys. Rev. B},
  volume = {100},
  issue = {13},
  pages = {134306},
  numpages = {26},
  year = {2019},
  month = {Oct},
  publisher = {American Physical Society},
  doi = {10.1103/PhysRevB.100.134306},
  url = {https://link.aps.org/doi/10.1103/PhysRevB.100.134306}
}

@article{jianMeasurementInducedCriticality2020,
  title = {Measurement-induced criticality in random quantum circuits},
  author = {Jian, Chao-Ming and You, Yi-Zhuang and Vasseur, Romain and Ludwig, Andreas W. W.},
  journal = {Phys. Rev. B},
  volume = {101},
  issue = {10},
  pages = {104302},
  numpages = {11},
  year = {2020},
  month = {Mar},
  publisher = {American Physical Society},
  doi = {10.1103/PhysRevB.101.104302},
  url = {https://link.aps.org/doi/10.1103/PhysRevB.101.104302}
}

@article{zabaloCriticalProperties2020,
  title = {Critical properties of the measurement-induced transition in random quantum circuits},
  author = {Zabalo, Aidan and Gullans, Michael J. and Wilson, Justin H. and Gopalakrishnan, Sarang and Huse, David A. and Pixley, J. H.},
  journal = {Phys. Rev. B},
  volume = {101},
  issue = {6},
  pages = {060301(R)},
  numpages = {5},
  year = {2020},
  month = {Feb},
  publisher = {American Physical Society},
  doi = {10.1103/PhysRevB.101.060301},
  url = {https://link.aps.org/doi/10.1103/PhysRevB.101.060301}
}

@article{fisher2023random,
  title={Random quantum circuits},
  author={Fisher, Matthew PA and Khemani, Vedika and Nahum, Adam and Vijay, Sagar},
  journal={Annual Review of Condensed Matter Physics},
  volume={14},
  number={1},
  pages={335--379},
  year={2023},
  publisher={Annual Reviews}
}

@article{gullansDynamicalPurification2020,
  title = {Dynamical Purification Phase Transition Induced by Quantum Measurements},
  author = {Gullans, Michael J. and Huse, David A.},
  journal = {Phys. Rev. X},
  volume = {10},
  issue = {4},
  pages = {041020},
  numpages = {28},
  year = {2020},
  month = {Oct},
  publisher = {American Physical Society},
  doi = {10.1103/PhysRevX.10.041020},
  url = {https://link.aps.org/doi/10.1103/PhysRevX.10.041020}
}

@article{noelMeasurementInducedQuantum2022,
  title={Measurement-induced quantum phases realized in a trapped-ion quantum computer},
  author={Noel, Crystal and Niroula, Pradeep and Zhu, Daiwei and Risinger, Andrew and Egan, Laird and Biswas, Debopriyo and Cetina, Marko and Gorshkov, Alexey V and Gullans, Michael J and Huse, David A and others},
  journal={Nature Physics},
  volume={18},
  number={7},
  pages={760--764},
  year={2022},
  publisher={Nature Publishing Group UK London}
}

@article{friedmanMeasurementInducedPhases2023,
  title = {Measurement-Induced Phases of Matter Require Feedback},
  author = {Friedman, Aaron J. and Hart, Oliver and Nandkishore, Rahul},
  journal = {PRX Quantum},
  volume = {4},
  issue = {4},
  pages = {040309},
  numpages = {46},
  year = {2023},
  month = {Oct},
  publisher = {American Physical Society},
  doi = {10.1103/PRXQuantum.4.040309},
  url = {https://link.aps.org/doi/10.1103/PRXQuantum.4.040309}
}

@article{MIPTdmrg2020,
  title = {Measurement-induced phase transition: A case study in the nonintegrable model by density-matrix renormalization group calculations},
  author = {Tang, Qicheng and Zhu, W.},
  journal = {Phys. Rev. Res.},
  volume = {2},
  issue = {1},
  pages = {013022},
  numpages = {7},
  year = {2020},
  month = {Jan},
  publisher = {American Physical Society},
  doi = {10.1103/PhysRevResearch.2.013022},
  url = {https://link.aps.org/doi/10.1103/PhysRevResearch.2.013022}
}

@article{MIPTscalable2020,
  title = {Scalable Probes of Measurement-Induced Criticality},
  author = {Gullans, Michael J. and Huse, David A.},
  journal = {Phys. Rev. Lett.},
  volume = {125},
  issue = {7},
  pages = {070606},
  numpages = {6},
  year = {2020},
  month = {Aug},
  publisher = {American Physical Society},
  doi = {10.1103/PhysRevLett.125.070606},
  url = {https://link.aps.org/doi/10.1103/PhysRevLett.125.070606}
}

@article{lavasani2021measurement,
  title={Measurement-induced topological entanglement transitions in symmetric random quantum circuits},
  author={Lavasani, Ali and Alavirad, Yahya and Barkeshli, Maissam},
  journal={Nature Physics},
  volume={17},
  number={3},
  pages={342--347},
  year={2021},
  publisher={Nature Publishing Group UK London}
}

@article{universalityMIPTPRL2020,
  title = {Universality of Entanglement Transitions from Stroboscopic to Continuous Measurements},
  author = {Szyniszewski, M. and Romito, A. and Schomerus, H.},
  journal = {Phys. Rev. Lett.},
  volume = {125},
  issue = {21},
  pages = {210602},
  numpages = {6},
  year = {2020},
  month = {Nov},
  publisher = {American Physical Society},
  doi = {10.1103/PhysRevLett.125.210602},
  url = {https://link.aps.org/doi/10.1103/PhysRevLett.125.210602}
}

@article{mblMIPT2020,
  title = {Measurement-induced entanglement transitions in many-body localized systems},
  author = {Lunt, Oliver and Pal, Arijeet},
  journal = {Phys. Rev. Res.},
  volume = {2},
  issue = {4},
  pages = {043072},
  numpages = {10},
  year = {2020},
  month = {Oct},
  publisher = {American Physical Society},
  doi = {10.1103/PhysRevResearch.2.043072},
  url = {https://link.aps.org/doi/10.1103/PhysRevResearch.2.043072}
}

@article{baoSymmetryEnriched2021,
title = {Symmetry enriched phases of quantum circuits},
journal = {Annals of Physics},
volume = {435},
pages = {168618},
year = {2021},
note = {Special issue on Philip W. Anderson},
issn = {0003-4916},
doi = {https://doi.org/10.1016/j.aop.2021.168618},
url = {https://www.sciencedirect.com/science/article/pii/S0003491621002244},
author = {Yimu Bao and Soonwon Choi and Ehud Altman},
}

@article{MIPTising2021,
  title = {Measurement-induced entanglement transitions in the quantum Ising chain: From infinite to zero clicks},
  author = {Turkeshi, Xhek and Biella, Alberto and Fazio, Rosario and Dalmonte, Marcello and Schir\'o, Marco},
  journal = {Phys. Rev. B},
  volume = {103},
  issue = {22},
  pages = {224210},
  numpages = {13},
  year = {2021},
  month = {Jun},
  publisher = {American Physical Society},
  doi = {10.1103/PhysRevB.103.224210},
  url = {https://link.aps.org/doi/10.1103/PhysRevB.103.224210}
}

@article{zabaloOperatorScaling2022,
  title = {Operator Scaling Dimensions and Multifractality at Measurement-Induced Transitions},
  author = {Zabalo, A. and Gullans, M. J. and Wilson, J. H. and Vasseur, R. and Ludwig, A. W. W. and Gopalakrishnan, S. and Huse, David A. and Pixley, J. H.},
  journal = {Phys. Rev. Lett.},
  volume = {128},
  issue = {5},
  pages = {050602},
  numpages = {6},
  year = {2022},
  month = {Feb},
  publisher = {American Physical Society},
  doi = {10.1103/PhysRevLett.128.050602},
  url = {https://link.aps.org/doi/10.1103/PhysRevLett.128.050602}
}

@article{mixedStateOrderMeasurementfeedback2023,
  title = {Mixed-State Long-Range Order and Criticality from Measurement and Feedback},
  author = {Lu, Tsung-Cheng and Zhang, Zhehao and Vijay, Sagar and Hsieh, Timothy H.},
  journal = {PRX Quantum},
  volume = {4},
  issue = {3},
  pages = {030318},
  numpages = {25},
  year = {2023},
  month = {Aug},
  publisher = {American Physical Society},
  doi = {10.1103/PRXQuantum.4.030318},
  url = {https://link.aps.org/doi/10.1103/PRXQuantum.4.030318}
}

@article{MIPTexpSC2023,
  title={Measurement-induced entanglement phase transition on a superconducting quantum processor with mid-circuit readout},
  author={Koh, Jin Ming and Sun, Shi-Ning and Motta, Mario and Minnich, Austin J},
  journal={Nature Physics},
  volume={19},
  number={9},
  pages={1314--1319},
  year={2023},
  publisher={Nature Publishing Group UK London}
}

@article{localizationMeas2023,
  title = {Localization properties in disordered quantum many-body dynamics under continuous measurement},
  author = {Yamamoto, Kazuki and Hamazaki, Ryusuke},
  journal = {Phys. Rev. B},
  volume = {107},
  issue = {22},
  pages = {L220201},
  numpages = {7},
  year = {2023},
  month = {Jun},
  publisher = {American Physical Society},
  doi = {10.1103/PhysRevB.107.L220201},
  url = {https://link.aps.org/doi/10.1103/PhysRevB.107.L220201}
}

@article{purificatoinUniversalPRX2025,
  title = {Universality Classes for Purification in Nonunitary Quantum Processes},
  author = {De Luca, Andrea and Liu, Chunxiao and Nahum, Adam and Zhou, Tianci},
  journal = {Phys. Rev. X},
  volume = {15},
  issue = {4},
  pages = {041024},
  numpages = {36},
  year = {2025},
  month = {Nov},
  publisher = {American Physical Society},
  doi = {10.1103/wlj6-mkk4},
  url = {https://link.aps.org/doi/10.1103/wlj6-mkk4}
}

@article{googleExp2023,
  title = {Measurement-Induced Entanglement and Teleportation on a Noisy Quantum Processor},
  author = {{Google Quantum AI and Collaborators} and Hoke, J. C. and Ippoliti, M. and Rosenberg, E. and Abanin, D. and Acharya, R. and Andersen, T. I. and Ansmann, M. and Arute, F. and Arya, K. and Asfaw, A. and Atalaya, J. and Bardin, J. C. and Bengtsson, A. and Bortoli, G. and Bourassa, A. and Bovaird, J. and Brill, L. and Broughton, M. and Buckley, B. B. and Buell, D. A. and Burger, T. and Burkett, B. and Bushnell, N. and Chen, Z. and Chiaro, B. and Chik, D. and Cogan, J. and Collins, R. and Conner, P. and Courtney, W. and Crook, A. L. and Curtin, B. and Dau, A. G. and Debroy, D. M. and Del Toro Barba, A. and Demura, S. and Di Paolo, A. and Drozdov, I. K. and Dunsworth, A. and Eppens, D. and Erickson, C. and Farhi, E. and Fatemi, R. and Ferreira, V. S. and Burgos, L. F. and Forati, E. and Fowler, A. G. and Foxen, B. and Giang, W. and Gidney, C. and Gilboa, D. and Giustina, M. and Gosula, R. and Gross, J. A. and Habegger, S. and Hamilton, M. C. and Hansen, M. and Harrigan, M. P. and Harrington, S. D. and Heu, P. and Hoffmann, M. R. and Hong, S. and Huang, T. and Huff, A. and Huggins, W. J. and Isakov, S. V. and Iveland, J. and Jeffrey, E. and Jiang, Z. and Jones, C. and Juhas, P. and Kafri, D. and Kechedzhi, K. and Khattar, T. and Khezri, M. and Kieferov{\'a}, M. and Kim, S. and Kitaev, A. and Klimov, P. V. and Klots, A. R. and Korotkov, A. N. and Kostritsa, F. and Kreikebaum, J. M. and Landhuis, D. and Laptev, P. and Lau, K.-M. and Laws, L. and Lee, J. and Lee, K. W. and Lensky, Y. D. and Lester, B. J. and Lill, A. T. and Liu, W. and Locharla, A. and Martin, O. and McClean, J. R. and McEwen, M. and Miao, K. C. and Mieszala, A. and Montazeri, S. and Morvan, A. and Movassagh, R. and Mruczkiewicz, W. and Neeley, M. and Neill, C. and Nersisyan, A. and Newman, M. and Ng, J. H. and Nguyen, A. and Nguyen, M. and Niu, M. Y. and O'Brien, T. E. and Omonije, S. and Opremcak, A. and Petukhov, A. and Potter, R. and Pryadko, L. P. and Quintana, C. and Rocque, C. and Rubin, N. C. and Saei, N. and Sank, D. and Sankaragomathi, K. and Satzinger, K. J. and Schurkus, H. F. and Schuster, C. and Shearn, M. J. and Shorter, A. and Shutty, N. and Shvarts, V. and Skruzny, J. and Smith, W. C. and Somma, R. and Sterling, G. and Strain, D. and Szalay, M. and Torres, A. and Vidal, G. and Villalonga, B. and Heidweiller, C. V. and White, T. and Woo, B. W. K. and Xing, C. and Yao, Z. J. and Yeh, P. and Yoo, J. and Young, G. and Zalcman, A. and Zhang, Y. and Zhu, N. and Zobrist, N. and Neven, H. and Babbush, R. and Bacon, D. and Boixo, S. and Hilton, J. and Lucero, E. and Megrant, A. and Kelly, J. and Chen, Y. and Smelyanskiy, V. and Mi, X. and Khemani, V. and Roushan, P.},
  year = 2023,
  month = oct,
  journal = {Nature},
  volume = {622},
  number = {7983},
  pages = {481--486},
  issn = {0028-0836, 1476-4687},
  doi = {10.1038/s41586-023-06505-7},
  langid = {english}
}

@article{topo2dMonitoredPRL2021,
  title = {Topological Order and Criticality in $(2+1)\mathrm{D}$ Monitored Random Quantum Circuits},
  author = {Lavasani, Ali and Alavirad, Yahya and Barkeshli, Maissam},
  journal = {Phys. Rev. Lett.},
  volume = {127},
  issue = {23},
  pages = {235701},
  numpages = {6},
  year = {2021},
  month = {Dec},
  publisher = {American Physical Society},
  doi = {10.1103/PhysRevLett.127.235701},
  url = {https://link.aps.org/doi/10.1103/PhysRevLett.127.235701}
}

@article{symmetrytopoMonitoredPRB2026,
  title = {Symmetry and topology of monitored quantum dynamics},
  author = {Xiao, Zhenyu and Kawabata, Kohei},
  journal = {Phys. Rev. B},
  volume = {113},
  issue = {13},
  pages = {134307},
  numpages = {16},
  year = {2026},
  month = {Apr},
  publisher = {American Physical Society},
  doi = {10.1103/1q83-jcbw},
  url = {https://link.aps.org/doi/10.1103/1q83-jcbw}
}

@article{choiQECPRL2020,
  title = {Quantum Error Correction in Scrambling Dynamics and Measurement-Induced Phase Transition},
  author = {Choi, Soonwon and Bao, Yimu and Qi, Xiao-Liang and Altman, Ehud},
  journal = {Phys. Rev. Lett.},
  volume = {125},
  issue = {3},
  pages = {030505},
  numpages = {6},
  year = {2020},
  month = {Jul},
  publisher = {American Physical Society},
  doi = {10.1103/PhysRevLett.125.030505},
  url = {https://link.aps.org/doi/10.1103/PhysRevLett.125.030505}
}

@article{qecRandomFanPRB2021,
  title = {Self-organized error correction in random unitary circuits with measurement},
  author = {Fan, Ruihua and Vijay, Sagar and Vishwanath, Ashvin and You, Yi-Zhuang},
  journal = {Phys. Rev. B},
  volume = {103},
  issue = {17},
  pages = {174309},
  numpages = {26},
  year = {2021},
  month = {May},
  publisher = {American Physical Society},
  doi = {10.1103/PhysRevB.103.174309},
  url = {https://link.aps.org/doi/10.1103/PhysRevB.103.174309}
}

@article{liFisherQECPRB2021,
  title = {Statistical mechanics of quantum error correcting codes},
  author = {Li, Yaodong and Fisher, Matthew P. A.},
  journal = {Phys. Rev. B},
  volume = {103},
  issue = {10},
  pages = {104306},
  numpages = {19},
  year = {2021},
  month = {Mar},
  publisher = {American Physical Society},
  doi = {10.1103/PhysRevB.103.104306},
  url = {https://link.aps.org/doi/10.1103/PhysRevB.103.104306}
}

@article{adaptiveExpfeig2023,
  title={Experimental demonstration of the advantage of adaptive quantum circuits (2023)},
  author={Foss-Feig, M and Tikku, A and Lu, TC and Mayer, K and Iqbal, M and Gatterman, TM and Gerber, JA and Gilmore, K and Gresh, D and Hankin, A and others},
  journal={arXiv preprint arXiv:2302.03029}
}

@article{EEprepZlatkoPRXqu2024,
  title = {Efficient Long-Range Entanglement Using Dynamic Circuits},
  author = {B\"aumer, Elisa and Tripathi, Vinay and Wang, Derek S. and Rall, Patrick and Chen, Edward H. and Majumder, Swarnadeep and Seif, Alireza and Minev, Zlatko K.},
  journal = {PRX Quantum},
  volume = {5},
  issue = {3},
  pages = {030339},
  numpages = {20},
  year = {2024},
  month = {Aug},
  publisher = {American Physical Society},
  doi = {10.1103/PRXQuantum.5.030339},
  url = {https://link.aps.org/doi/10.1103/PRXQuantum.5.030339}
}

@article{EEnegativityCriticalityFisher2021,
  title = {Entanglement Negativity at Measurement-Induced Criticality},
  author = {Sang, Shengqi and Li, Yaodong and Zhou, Tianci and Chen, Xiao and Hsieh, Timothy H. and Fisher, Matthew P.A.},
  journal = {PRX Quantum},
  volume = {2},
  issue = {3},
  pages = {030313},
  numpages = {23},
  year = {2021},
  month = {Jul},
  publisher = {American Physical Society},
  doi = {10.1103/PRXQuantum.2.030313},
  url = {https://link.aps.org/doi/10.1103/PRXQuantum.2.030313}
}

@article{khemaniMeasOnly2021,
  title = {Entanglement Phase Transitions in Measurement-Only Dynamics},
  author = {Ippoliti, Matteo and Gullans, Michael J. and Gopalakrishnan, Sarang and Huse, David A. and Khemani, Vedika},
  journal = {Phys. Rev. X},
  volume = {11},
  issue = {1},
  pages = {011030},
  numpages = {23},
  year = {2021},
  month = {Feb},
  publisher = {American Physical Society},
  doi = {10.1103/PhysRevX.11.011030},
  url = {https://link.aps.org/doi/10.1103/PhysRevX.11.011030}
}

@article{measOnlyToric2024,
  title = {Measurement-only dynamical phase transition of topological and boundary order in toric code and gauge Higgs models},
  author = {Orito, Takahiro and Kuno, Yoshihito and Ichinose, Ikuo},
  journal = {Phys. Rev. B},
  volume = {109},
  issue = {22},
  pages = {224306},
  numpages = {10},
  year = {2024},
  month = {Jun},
  publisher = {American Physical Society},
  doi = {10.1103/PhysRevB.109.224306},
  url = {https://link.aps.org/doi/10.1103/PhysRevB.109.224306}
}

@article{measOnlySpinLiquidVijayPRB2023,
  title = {Monitored quantum dynamics and the {K}itaev spin liquid},
  author = {Lavasani, Ali and Luo, Zhu-Xi and Vijay, Sagar},
  journal = {Phys. Rev. B},
  volume = {108},
  issue = {11},
  pages = {115135},
  numpages = {26},
  year = {2023},
  month = {Sep},
  publisher = {American Physical Society},
  doi = {10.1103/PhysRevB.108.115135},
  url = {https://link.aps.org/doi/10.1103/PhysRevB.108.115135}
}

@article{sangMeasOnlyPRR2021,
  title = {Measurement-protected quantum phases},
  author = {Sang, Shengqi and Hsieh, Timothy H.},
  journal = {Phys. Rev. Res.},
  volume = {3},
  issue = {2},
  pages = {023200},
  numpages = {9},
  year = {2021},
  month = {Jun},
  publisher = {American Physical Society},
  doi = {10.1103/PhysRevResearch.3.023200},
  url = {https://link.aps.org/doi/10.1103/PhysRevResearch.3.023200}
}

@article{loopMeasOnlyPRX2023,
  title = {Majorana Loop Models for Measurement-Only Quantum Circuits},
  author = {Klocke, Kai and Buchhold, Michael},
  journal = {Phys. Rev. X},
  volume = {13},
  issue = {4},
  pages = {041028},
  numpages = {33},
  year = {2023},
  month = {Nov},
  publisher = {American Physical Society},
  doi = {10.1103/PhysRevX.13.041028},
  url = {https://link.aps.org/doi/10.1103/PhysRevX.13.041028}
}

@article{SPTMeasOnly2022,
  title={Emergence symmetry protected topological phase in spatially tuned measurement-only circuit},
  author={Kuno, Yoshihito and Ichinose, Ikuo},
  journal={arXiv preprint arXiv:2212.13142},
  year={2022}
}

@article{yuclusterIsingMeasOnly2025,
  title={Measurement-driven transitions between area law phases},
  author={Yu, Hui and Hu, Jiangping},
  journal={Physica Scripta},
  volume={100},
  number={6},
  pages={065950},
  year={2025},
  publisher={IOP Publishing}
}

@article{klockeTopologicalOrderEntanglement2022,
  title = {Topological Order and Entanglement Dynamics in the Measurement-Only {{XZZX}} Quantum Code},
  author = {Klocke, Kai and Buchhold, Michael},
  year = 2022,
  month = sep,
  journal = {Physical Review B},
  volume = {106},
  number = {10},
  pages = {104307},
  publisher = {American Physical Society},
  doi = {10.1103/PhysRevB.106.104307}
}

@article{kunoProductionLatticeGauge2023,
  title = {Production of Lattice Gauge {{Higgs}} Topological States in a Measurement-Only Quantum Circuit},
  author = {Kuno, Yoshihito and Ichinose, Ikuo},
  year = 2023,
  month = jun,
  journal = {Physical Review B},
  volume = {107},
  number = {22},
  pages = {224305},
  publisher = {American Physical Society},
  doi = {10.1103/PhysRevB.107.224305}
}

@article{langEntanglementTransitionProjective2020,
  title = {Entanglement Transition in the Projective Transverse Field {{Ising}} Model},
  author = {Lang, Nicolai and B{\"u}chler, Hans Peter},
  year = 2020,
  month = sep,
  journal = {Physical Review B},
  volume = {102},
  number = {9},
  pages = {094204},
  issn = {2469-9950, 2469-9969},
  doi = {10.1103/PhysRevB.102.094204},
  langid = {english}
}

@article{negativityLu2020,
  title={Entanglement negativity at the critical point of measurement-driven transition},
  author={Shi, Bowen and Dai, Xin and Lu, Yuan-Ming},
  journal={arXiv preprint arXiv:2012.00040},
  year={2020}
}

@article{aaronsonImprovedSimulationStabilizer2004,
  title = {Improved Simulation of Stabilizer Circuits},
  author = {Aaronson, Scott and Gottesman, Daniel},
  year = 2004,
  month = nov,
  journal = {Physical Review A},
  volume = {70},
  number = {5},
  pages = {052328},
  publisher = {American Physical Society},
  doi = {10.1103/PhysRevA.70.052328}
}

@article{gottesmanClassQuantumErrorcorrecting1996,
  title = {Class of Quantum Error-Correcting Codes Saturating the Quantum {{Hamming}} Bound},
  author = {Gottesman, Daniel},
  year = 1996,
  month = sep,
  journal = {Physical Review A},
  volume = {54},
  number = {3},
  pages = {1862--1868},
  publisher = {American Physical Society},
  doi = {10.1103/PhysRevA.54.1862}
}

@misc{gottesmanHeisenbergRepresentationQuantum1998,
  title = {The {{Heisenberg Representation}} of {{Quantum Computers}}},
  author = {Gottesman, Daniel},
  year = 1998,
  month = jul,
  number = {arXiv:quant-ph/9807006},
  eprint = {quant-ph/9807006},
  publisher = {arXiv},
  doi = {10.48550/arXiv.quant-ph/9807006},
  archiveprefix = {arXiv}
}

@article{liQuantumZenoEffect2018,
  title = {Quantum {{Zeno}} Effect and the Many-Body Entanglement Transition},
  author = {Li, Yaodong and Chen, Xiao and Fisher, Matthew P. A.},
  year = 2018,
  month = nov,
  journal = {Physical Review B},
  volume = {98},
  number = {20},
  pages = {205136},
  issn = {2469-9950, 2469-9969},
  doi = {10.1103/PhysRevB.98.205136},
  langid = {english}
}

@article{morral-yepesDetectingStabilizingMeasurementinduced2023,
  title = {Detecting and Stabilizing Measurement-Induced Symmetry-Protected Topological Phases in Generalized Cluster Models},
  author = {{Morral-Yepes}, Ra{\'u}l and Pollmann, Frank and Lovas, Izabella},
  year = 2023,
  month = dec,
  journal = {Physical Review B},
  volume = {108},
  number = {22},
  pages = {224304},
  issn = {2469-9950, 2469-9969},
  doi = {10.1103/PhysRevB.108.224304},
  langid = {english}
}

@article{sangMeasurementprotectedQuantumPhases2021,
  title = {Measurement-Protected Quantum Phases},
  author = {Sang, Shengqi and Hsieh, Timothy H.},
  year = 2021,
  month = jun,
  journal = {Physical Review Research},
  volume = {3},
  number = {2},
  pages = {023200},
  issn = {2643-1564},
  doi = {10.1103/PhysRevResearch.3.023200},
  langid = {english}
}

@misc{yuGaplessSymmetryProtectedTopological2025,
  title = {Gapless {{Symmetry-Protected Topological States}} in {{Measurement-Only Circuits}}},
  author = {Yu, Xue-Jia and Yang, Sheng and Liu, Shuo and Lin, Hai-Qing and Jian, Shao-Kai},
  year = 2025,
  month = jan,
  number = {arXiv:2501.03851},
  eprint = {2501.03851},
  primaryclass = {cond-mat},
  publisher = {arXiv},
  doi = {10.48550/arXiv.2501.03851},
  archiveprefix = {arXiv}
}

@article{cuiLocalCharacterizationOnedimensional2013,
  title = {Local Characterization of One-Dimensional Topologically Ordered States},
  author = {Cui, Jian and Amico, Luigi and Fan, Heng and Gu, Mile and Hamma, Alioscia and Vedral, Vlatko},
  year = 2013,
  month = sep,
  journal = {Physical Review B},
  volume = {88},
  number = {12},
  pages = {125117},
  issn = {1098-0121, 1550-235X},
  doi = {10.1103/PhysRevB.88.125117},
  urldate = {2025-11-24},
  copyright = {http://link.aps.org/licenses/aps-default-license},
  langid = {english}
}

@article{pollmannSymmetryProtectionTopological2012,
  title = {Symmetry protection of topological phases in one-dimensional quantum spin systems},
  author = {Pollmann, Frank and Berg, Erez and Turner, Ari M. and Oshikawa, Masaki},
  journal = {Phys. Rev. B},
  volume = {85},
  issue = {7},
  pages = {075125},
  numpages = {9},
  year = {2012},
  month = {Feb},
  publisher = {American Physical Society},
  doi = {10.1103/PhysRevB.85.075125},
  url = {https://link.aps.org/doi/10.1103/PhysRevB.85.075125}
}

@article{scaffidiGaplessSymmetryProtectedTopological2017,
  title = {Gapless {{Symmetry-Protected Topological Order}}},
  author = {Scaffidi, Thomas and Parker, Daniel E. and Vasseur, Romain},
  year = 2017,
  month = nov,
  journal = {Physical Review X},
  volume = {7},
  number = {4},
  pages = {041048},
  issn = {2160-3308},
  doi = {10.1103/PhysRevX.7.041048},
  urldate = {2026-02-07},
  copyright = {https://creativecommons.org/licenses/by/4.0/},
  langid = {english}
}
\end{document}